\documentclass[aps,prl,amsmath,twocolumn,showpacs]{revtex4}
\usepackage{graphicx}
\begin{document}
\title{New class of self-similar solutions for vacuum plasma expansion
\\
admitting monoenergetic ion spectra}
\author{Naveen Kumar\footnote{Electronic mail: kumar@tp1.uni-duesseldorf.de}}
\author{Alexander Pukhov}
\affiliation{Institut f\"ur Theoretische Physik I, Heinrich-Heine-Universit\"at D\"usseldorf, D-40225, Germany}

\begin{abstract}
We report a new class of self-similar solutions for plasma expanding into
vacuum that allows for quasi-monoenergetic ion spectra. A simple
analytical model takes into account externally controlled time-dependent
temperature of the hot electrons. When the laser temporal profile is
tailored properly, the quasi-neutral self-similar expansion of the
plasma results in ion concentration in the phase-space at a particular
velocity thus producing a quasi-monoenergetic spectrum. We prove this
analytical prediction using a 1D partice-in-cell (PIC) simulation
where the time-dependent plasma temperature is controlled by two laser
pulses shot at a foil with a suitable time delay between them.  
\end{abstract}
\maketitle

Generation of highly energetic ions and protons beams from the laser interaction
with thin foil targets in relativistic regime
($I\,\lambda^{2}>10^{18}$ W cm$^{-2}\,\mu$m$^2$, where $I$ is the
intensity of the laser and $\lambda$ its wavelength) is one of the
fast developing research fields \cite{hatchett00,snavely00,wilks01,
hagelich02,pukhov01,habs01}. Particularly interesting is the
generation of quasi-monoenergetic ion beams, because these 
have a number of important potential applications in
medical physics, inertial confinement fusion, compact
ion accelerators \cite{hagelich06,sch06,avetisyan06,tikhonchuk}. 

When the laser pulse strikes a thin solid density foil, it heats electrons
at the front surface of the target. These hot electrons traverse
the target and leave it at the rear side. A space charge is build up
that leads to a huge electrostatic potential. This, in turn,
accelerates the background ions to high energies.  This mechanism is known
as target normal sheath acceleration (TNSA) \cite{wilks01,pukhov01,hagelich02,habs01,hagelich06,sch06,avetisyan06, tikhonchuk}. From theoretical point of view the generation of
energetic ions from thin foil targets has been studied well in past
based on the approach of self-similar expansion of
plasma\cite{gurevich66, allen70,murakami05}. In this model, the plasma
fluid is quasi-neutral, ions and electrons both have similar flow on
the ion sound time scale. In the works  \cite{mora03,murakami06}
also kinetic effects at the front of the expanding plasma were
included in this self-similar model for different geometries.

All the self-similar models up to now have predicted broad
band ion spectra, decaying exponentially at high energies. Yet,
most of practical applications require monoenergetic ion beams. A
straightforward way to overcome this difficulty is to select a group
of ions with the same energy, e.g. using ions with the same initial
conditions. Indeed, when a monolayer of protons, or a proton-rich dot
is placed on the back  surface of a higher-Z target, then all the light
protons are accelerated  nearly to the same energy
\cite{hagelich06,sch06}, because all the protons in 
the monolayer feel same electric field. Still, the number of
accelerated protons out of the monolayer and thus the efficiency of the
process are very limited. Recently, it was also predicted that ion
acceleration from multi-species targets may result in peaked spectra
\cite{avetisyan06,tikhonchuk}.

In this Letter, we report a new type of self-similar solution for
the quasi-neutral vacuum-plasma expansion that does allow for monoenergetic spikes in
the ion spectrum. As we will see, this new self-similar solution
requires a suitable time-dependent temperature of laser-generated hot
electrons. Experimentally, this corresponds to a tailored
temporal profile of the driving laser pulse, because the temperature
of the hot electrons is a definite function of the laser intensity
\cite{wilks92,pukhov1999}.  The rising electron temperature with time is
accompanied by a rising electrostatic field inside the expanding
plasma. Consequently, this results in a higher acceleration rate of
ions at 
later times. These rear ions catch up with the leading ions, which
began to accelerate earlier. When this catching up occurs, a bunch of
ions concentrated in phase space is formed. If driven too harsch, this
catching up may result in 
``{\em wave breaking}'' \cite{mora06}. However, when the temperature is
changed gently, one may stay ``just at the verge'' of
wavebreaking. This is exactly the solution we present here.

Under the quasi-neutral assumption \emph{i.e} $n_{e}\approx Z n_{i}$
where $n_{e}\, \text{and}\, n_{i}$ are the electron and ion densities
repectively, and $Z$ is the ionization state, the dynamics of the
plasma expansion is governed by the continuity equation

\begin{equation}\label{eq1} 
\frac{\partial n_{i}}{\partial t}+\frac{\partial}{\partial
r}(n_{i}\upsilon_{i})=0, 
\end{equation}

\noindent and the Euler equation for the ion fluid

\begin{equation}\label{eq2}
\frac{\partial \upsilon_{i}}{\partial t}+\upsilon_{i}\frac{\partial
\upsilon_{i}}{\partial r}+\frac{Z}{n_{i} m_{i}}\frac{\partial (n_{i}
T_{e}(t))}{\partial r}=0. 
\end{equation}

\noindent
Here $\upsilon_{i}$ is the ion fluid velocity, $m_{i}$ is the ion
mass, $T_{e}(t)=T_{0}(t)(n_{e}(0,t)/n_{e}(0,0))^{\gamma-1}$ is the
local electron temperature and $\gamma$ is the expansion polytrope
index for electrons. We also assume that the plasma electrons obey
the Boltzmann statistics  

\begin{equation}
n_{e}(r, t)=n_{e}(t) \text{exp}(e\phi/T_{e}),
\end{equation}

\noindent
where $n_{e}(t)$ is the time-dependent electron density, $-e$ is the
electron charge, and $\phi (r, t)$ is the electrostatic 
potential. The ions are assumed to be cold initially. 

It is well known that the above system of equations permits the
following self-similar solution \cite{murakami05} 

\begin{equation*}
\upsilon_{i(e)}=\dot{R}\,\xi,\;\xi=x/R,
\end{equation*}
\begin{equation}\label{old}
n_{e}\approx Z n_{i} = n_{00}(R_{0}/R)\; N_{i}(\xi),\; N_{i}(0)=1,
\end{equation}

\noindent
where $R(t)$ is the time dependent charactersitic size of the plasma, $\xi$ is the similarity coordinate ($x$ is the position
coordinate), and $N_{i}$ is a positive unknown function. The subscripts $i$ and $e$ refer to the ions and
electrons respectively. The rationale behind this self-similar
solution is that a linear velocity-radius relation is the correct
limit for the asymptotic stage of the expansion, whenever the $R(t)$
greatly exceeds the initial value $R_{0}=R(0)$ \cite{murakami05}. We may
see that this solution rules out the possibility of generation of
quasi-monoenergetic ion beams because the ion velocity scales linearly
with the similarity coordinate $\xi$. 

In this Letter, we introduce a more general self-similarity ansatz
and assume that the plasma expansion scales as

\begin{equation*}
\upsilon_{i(e)}=\dot{R}\,f(\xi),\;\xi=x/R,
\end{equation*}
\begin{equation}\label{eq3}
n_{e}\approx Z n_{i}=n_{00}(R_{0}/R)\; N_{i}(\xi),\; N_{i}(0)=1,
\end{equation}

\noindent
where $f(\xi)$ is a function which is yet to be determined.
Because the new solution must satisfy the equation of continuity,
we have from Eq.\eqref{eq1} 

\begin{equation}
f(\xi)=\xi-C/N_{i}(\xi). \label{C}
\end{equation}

\noindent Here $C$ is a constant of integration. Physically, the fluid
velocity $\upsilon_i$ must be monotonic. Consequently, the solution
\eqref{C} has a meaning only in the range $\xi < \xi_{\max}$
where $\xi_{\max}$ marks the front end of the plasma expansion in
vacuum. We require that $C < \xi_{\max}
N_{i}\left(\xi_{\max}\right)$. This restriction means that $C$ is
 small, because $N_i$ decays exponentially with $\xi$. Thus, the new
solution is hardly distinguishable from the classic one at small
$\xi$. Only approaching $\xi_{\max}$ the solution is modified
significantly. For the sake of accuracy, we mention that our solution
leads to a non-zero fluid velocity at the origin, $\upsilon_i(0) = -C
\dot{R}$. This velocity, however, is exponentially small (just as $C$)
and has no great physical meaning. It is well known, that self-similar
solutions are not well defined at the origin and must be considered at
some distance from the origin only \cite{gurevich66}. 

One easily retrieves the old similarity ansatz \eqref{old} by 
setting $C=0$ in \eqref{C}. Thus, all the previous self-similar
solutions for plasma expansion into vacuum were missing the second
term in \eqref{C}. Yet, this term is very important. It is the second
term in \eqref{C} with the non-zero constant $C$ that allows for
a new class of self-similar solutions and gives the option for
quasi-monoenergetic ion spectra. 

The condition to have a monoenergetic spike in the ion spectrum is the
presence of a stationary point in the expansion velocity, 
$d\upsilon_i (\xi_\text{m})/d\xi = 0$,  that gives

\begin{equation}
C \frac{d N_{i}\left(\xi_\text{m}\right)}{d \xi} =
-N_{i}^2\left(\xi_\text{m}\right).
\label{mono_condition}
\end{equation}

\noindent The condition \eqref{mono_condition} means that there is a
group of ions around the point $\xi_\text{m}$ which move all with the
same velocity 

\begin{equation}
\upsilon_\text{m} = \dot{R} f\left(\xi_\text{m}\right).
\label{mono_velocity}
\end{equation}

\noindent This ion beam produces a spike in the energy spectrum around
$\varepsilon_\text{m} = m_i \upsilon_\text{m}^2/2$.

Now we substitute the formula \eqref{C} in Eq.\eqref{eq2} and obtain

\begin{equation}\label{eq4}
\frac{d N_{i}}{d \xi}=\frac{N_{i}^2[ N_{i}\xi-C(1+a)]}{[ a C^2-b
N_{i}^{\gamma+1}]}, 
\end{equation}

\noindent
where we have used the following relations

\begin{equation}\label{acc}
 \ddot{R}=\frac {\dot{R}^2} {a R}=\frac {Z \gamma}{b m_{i} R_{0}}\Big(\frac{R_{0}}{R}\Big)^\gamma T_{0}(t),
 \end{equation}

\noindent
  $a\, \text{and}\, b$ are constants. It follows from Eqs.~\eqref{mono_condition} and \eqref{eq4} that the
 density of the monoenergetic  ion beam is (in the isothermal case, $\gamma=1$)

\begin{equation}\label{mono_density}
N_{i}\left(\xi_\text{m}\right) =\left(C/2b\right) \left[ 
\xi_\text{m} + \left(\xi_\text{m} ^2-4b\right)^{1/2}\right],
\end{equation}
\noindent and its velocity
\begin{equation}\label{vmono}
\upsilon_\text{m} = \dot{R} \left( \xi_\text{m} -
2 b\left(\xi_\text{m} + \left(\xi_\text{m} ^2-4b\right)^{-1/2}\right)\right).
\end{equation}
\noindent 
It is clear that the density and energy of the monoenergetic bunch do not depend on the constant $a$. Hence, in principle this constant can be chosen freely. The constant $b$ can be expressed in terms of the constant $C$ and density of the monoenergetic bunch, which is sought experimentally, as
\begin{equation}
 b = C \left(N_{i}\left(\xi_\text{m}\right)\xi_\text{m}-C\right) /  N_{i}^2 \left(\xi_\text{m}\right).
 \end{equation}


One can draw further conclusions
 regarding ion acceleration from the self-similar
 solution. We see from the Eq.\eqref{acc} that the ion acceleration is inversely proportional
to $R$, which is an increasing function of time. We obtain the expression for $R$
on integrating the first relation in Eq.\eqref{acc}
\begin{equation}\label{eq5}
\frac{R}{R_{0}}=\left(1+\frac{(a-1)}{a}\frac{\upsilon_{f}^{'}}{R_{0} } t\right)^{\frac{a}{a-1}},
\end{equation}

\noindent where $\upsilon_{f}^{'} = \upsilon_{f}\left(R_{0}/R_{f}\right)^{1/a}$. Here $\upsilon_{f}$ and $R_{f}$ are the
velocity and size of the expanding plasma, in the case of a
time independent electron temperature, respectively. These can be
determined from the results of Ref.\cite{murakami05}. Once the
expression for $R$ is obtained, we can determine the required electron
temperature profile from the second part of the Eq.~\eqref{acc}:
\begin{equation}\label{temp}
T_{0}(t)=\frac{b\, T_{f}}{Z \gamma\,
a}
\left(1+\frac{(a-1)}{a}\frac{\upsilon_{f}^{'}}{R_{0}}
t\right)^{\frac{2+a(\gamma-1)}{a-1}},
\end{equation}

\noindent where $T_{f}=m_{i}\upsilon_{f}^{' 2}$  is the time independent electron temperature. This relation describes the electron  temperature variation
in terms of the constants $a$ and $b$. These constants define how time
dependent electron temperature profile must be tailored to produce the
monoenergetic ion beam. The idea of controlling the plasma electron temperature has been recently studied by means of Vlasov and PIC simulations \cite{alex07}. It is found that a suitable temperature variation of the plasma electrons indeed can produce the mono-energetic ions.  

\begin{figure}
\includegraphics[height=0.42\textwidth,width=0.52\textwidth]{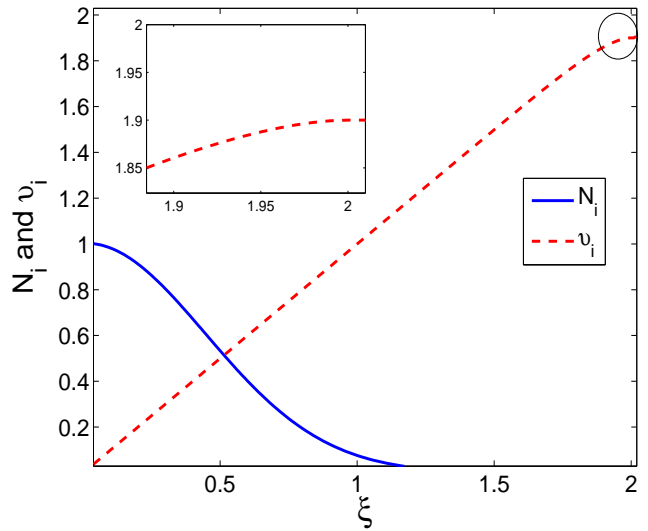}
\caption{Variation of $N_{i}\,\text{and}\,\upsilon_{i}$ with $\xi$
showing the self-similar flow and velocity of the expanding plasma. The inset is a zoomed in view of the velocity of expansion near the flattening point. The parameters are given in the text.}\label{fig1} 
\end{figure}

\begin{figure}
\includegraphics[height=0.42\textwidth,width=0.52\textwidth]{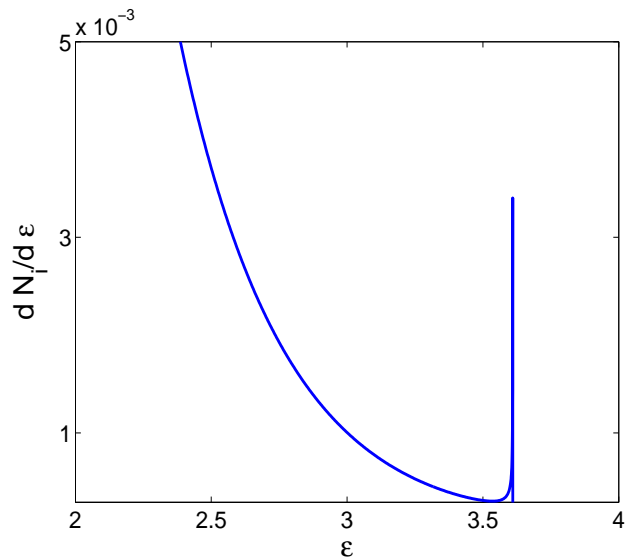}
\caption{Ion energy spectrum corresponding to the self-similar
solution given by Eq.\eqref{eq4} for $\gamma=1$. Here the energy is
normalized by $m_{i} \dot{R}^2/2$ and the ion density $N_{i}$ is
normalized by the value at the center of the thin foil
target. The other parameters are same as in the Fig.\ref{fig1}}\label{fig2} 
\end{figure}

An analytical solution of Eq.~\eqref{eq4} is hardly possible and
 we integrate it numerically. Fig.\ref{fig1} shows $N_{i}$
 and $\upsilon_{i}$ as a function of $\xi$ for the parameter set $a=5,
 b=0.19, C=3.3636\times10^{-6}, \text{and}\, \gamma=1$. Fig.\ref{fig1} describes the 
 self-similar expansion of the quasi-neutral plasma. The most
 important result is the behaviour of $\upsilon_{i}$ showing 
 flattening with $\xi$ near $\xi=2$. This flattening
manifests the production of monoenergetic ions. The velocity of
 expansion reaches its maximum at the flattening point. It corresponds
 also to the position of the ion front, where the self-similar
 solution has to be terminated. The energy spectrum of ions
produced by this self-similar solution is given in Fig.\ref{fig2}. 
We observe the quasi-monoenergetic bunch formation from the figure.  


We have performed a $1$D particle-in-cell (PIC) simulation of the
quasi-monoenergetic protons generation from a thin foil target  with
the help of  virtual laser plasma laboratory (VLPL) code
\cite{pukhov99}. In the simulation, we take a thin foil of thickness
$1.5 \lambda$ with an initial density ramp and solid density $30
n_{c}$, where $n_{c}$ is the critical density and $\lambda$ is the
wavelength of  the laser pulse. 

To tailor the time dependent temperature of the hot electrons, we use
two driving laser pulses shot at the target with a time delay between them. 
The simulation parameters for the
laser pulses are; $a_{1}=e A_{1}/mc^2=1,\,c\tau_{1}=5\lambda,\,
a_{2}=e A_{2}/mc^2 = 3,\,c\tau_{2}=15\lambda\;\text{and}\; \lambda=0.8\, \mu$m, where
$\tau_{1}$ and $\tau_{2}$ are the  pulse durations, and $a_{1,2}$ are the normalized vector potentials of the two laser pulses. Here subscripts
$1$ and $2$ represent the first and second laser pulse
respectively. The time delay between the two pulses is $60$~laser
periods. Both laser are $\hat{y}$-polarized and move in $\hat{x}$ direction.
The resolution in $x$-direction is $0.01\lambda$
while $200$ numerical particles per cells are chosen to run the
simulation. 

\begin{figure}
\includegraphics[height=0.62\textwidth,width=0.52\textwidth]{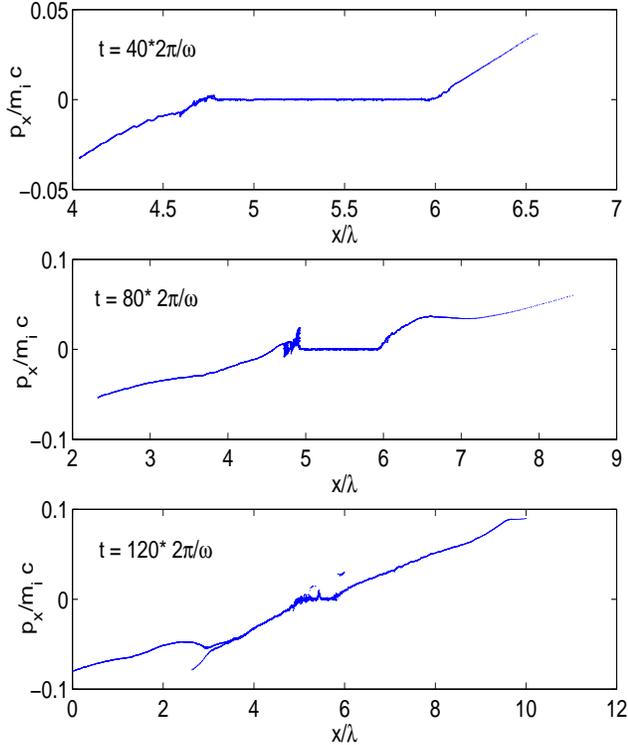}
\caption{The phase space evolution of  protons at the various stages of  thin foil expansion. The middle and lower panels of the figure describe the momenta flattening, which ultimately leads to the generation of monoenergetic protons from the thin foil target.}\label{fig3}
\end{figure}
\begin{figure}[!htp]
\includegraphics[height=0.45\textwidth,width=0.52\textwidth]{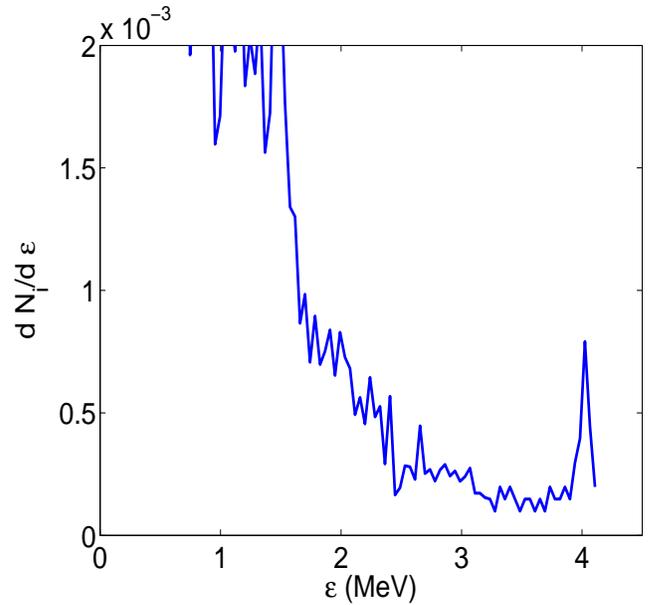}
\caption{The energy spectrum of the protons showing the monoenergetic
peak around  $4$ MeV from the PIC simulation. The proton density
$N_{i}$ is normalized by the critical density.}\label{fig5} 
\end{figure}

Fig.\ref{fig3} depicts the phase space evolution of protons at  the various
stages of plasma expansion. The upper panel depicts the stage of
the thin foil expansion under the influence of the first laser
pulse. It is clear that the single laser pulse leads to 
the standard self-similar plasma expansion given by  Eq.\eqref{old} and thus
clearly rules out the possibility of concentration of protons in the 
phase-space. The proton velocity scales linearly with the distance from the
target. Later (at $t=80\cdot 2\pi/ \omega$ where $\omega$ is the
carrier frequency of the laser pulse), when  the second pulse hits the
target (c.f. middle panel of the Fig.\ref{fig3}), the dynamics of the
expanding plasma changes and one can see momenta flattening in the phase
space. The momenta flattening forms 
the monoenergetic protons bunch. Physically, the second laser pulse
increases the hot electron temperature that leads to a rising
electrostatic field and a larger acceleration of ions at later
times. These ions, which form a second 
group,  catch up with the first group of ions, which began to
accelerate earlier. This process flattens the velocity profile and
forms a bunch of quasi-monoenergetic ions . As the
time passes, this bunch of ions accelerates and acquires
substantial energy as shown in the 
lower panel of the Fig.\ref{fig3}. Fig.\ref{fig5} 
shows the ion energy spectrum obtained in the PIC simulation. It
can be seen that there is a significant 
number of ions with the peak energy around $4$ MeV and the energy
spread is a few per cent. The  number density  of quasi-monoenergetic
ions can be further enhanced by  using higher density targets and
longer laser pulses. Using the two-pulse technique - or, more general,
a tailored time-dependent laser intensity - it will be possible to
produce light as well as heavy ions ($Z>1$) with monoenergetic
spectra. However, the heavy ion case may demand higher laser 
intensities.
 
In summary, we have found a new class of self-similar solutions for
the quasi-neutral vacuum-plasma expansion  that allows for quasi-monoenergetic ion
spectra. The quasi-monoenergetic spectrum corresponds to the flattening of
the ion momentum dependence on the expansion coordinate. Thus, ions
pile up at a particular velocity  in the phase space and form a quasi-monoenergetic bunch. This new scheme
demands for laser pulses with a tailored temporal profile. To prove
realizability of the new scheme, we have performed a 1D PIC simulation
using the code VLPL. In the PIC simulation, we use two laser pulses
hitting the foil target with a sutiable time delay between them.
Although the developed model is simple, it catches the essence
of the physical process quite well. The proposed scheme works already
for moderate laser powers which should make the laser-plasma generation of
monoenergetic protons (ions) a stable and well controlled process.

\acknowledgments This work was supported by the DFG through project TR-18.

\end{document}